\documentclass[a4paper]{panl}
\usepackage{cite}
\usepackage{wrapfig}
\usepackage{graphicx}
\usepackage{amssymb}
\usepackage{amsfonts}
\usepackage{amsmath}
\usepackage{longtable}
\usepackage{rotating}
\usepackage{lscape}
\usepackage{epsfig}
\usepackage{multirow}

\usepackage{lineno}

\originalTeX
\begin{document}

\title{Bayesian Approach to Particles Identification in the MPD Experiment}
\maketitle
\authors{V.A.\,Babkin$^{a,}$\footnote{E-mail: babkin@jinr.ru}, V.M.\,Baryshnikov$^{a}$, M.G.\,Buryakov$^{a}$, A.S.\,Burdyko$^{a}$,} 
\authors{S.G.\,Buzin$^{a}$, A.V.\,Dmitriev$^{a}$, V.I.\,Dronik$^{a,b}$, P.O.\,Dulov$^{a,c}$,}
\authors{A.A.\,Fedyunin$^{a}$, V.M.\,Golovatyuk$^{a}$, E.Yu.\,Kidanova$^{b}$, S.P.\,Lobastov$^{a}$,}
\authors{A.D.\,Pyatigor$^{b}$, M.M.\,Rumyantsev$^{a}$, K.A.\,Vokhmyanina$^{b}$}
\setcounter{footnote}{0}
\from{$^{a}$\,Joint Institute for Nucleart Research, Dubna, Russia}
\from{$^{b}$\,Belgorod National Research University, Belgorod, Russia}
\from{$^{c}$\,Plovdiv University ``Paisii Hilendarski'', Plovdiv, Bulgaria}


\begin{abstract}
Identification of particles generated by ion collisions in the NICA collider is one of the basic functions of the Multipurpose Detector (MPD). The main 
means of identification in MPD are the time-of-flight system (TOF) and the time-projection chamber (TPC). The article considers the optimization 
of the algorithms of particles identification by these systems. Under certain conditions, the use of the statistical Bayesian approach 
has made it possible to achieve an optimal ratio of the efficiency of particle identification and contamination by incorrectly defined particles.
\end{abstract}
\vspace*{6pt}


\label{sec:intro}
\section{Introduction}

The MPD experiment at NICA accelerator complex \cite{1:NICA-Kekelidze} at JINR (Dubna) is aimed to study of dense nuclear matter in the range of collision 
energies $\sqrt{s_{NN}} =$~4--11~GeV. The physical program of the NICA with heavy ions includes the following major goals: studying in-medium properties of 
hadrons and the nuclear matter equation of state, chiral symmetry phase transitions and search for signals of deconfinement and the QCD critical end point. 
To achieve these goals, it will be necessary to measure observables that are sensitive to high-density effects and phase transitions. That are, 
particle yields, particle type ratios, correlations and fluctuations. 

MPD \cite{2:MPD-EPJ} is the main experiment studying heavy ion collisions at the NICA accelerator complex. 
The MPD setup should be capable high-precision tracking, identification of particles in a large phase space, and very accurate measurements 
of centrality and event plane reconstruction.

The identification of charged hadrons in MPD at the first stage of experiment is carried out by two detector subsystems: the Time Projection 
Chamber (TPC) by energy losses and the Time-of-Flight system (TOF).

\label{sec:PID_algorithms}
\section{Identification of charged particles in MPD}

\begin{figure}[b]
\begin{center}
\vspace{-3mm}
\includegraphics[width=0.49\linewidth]{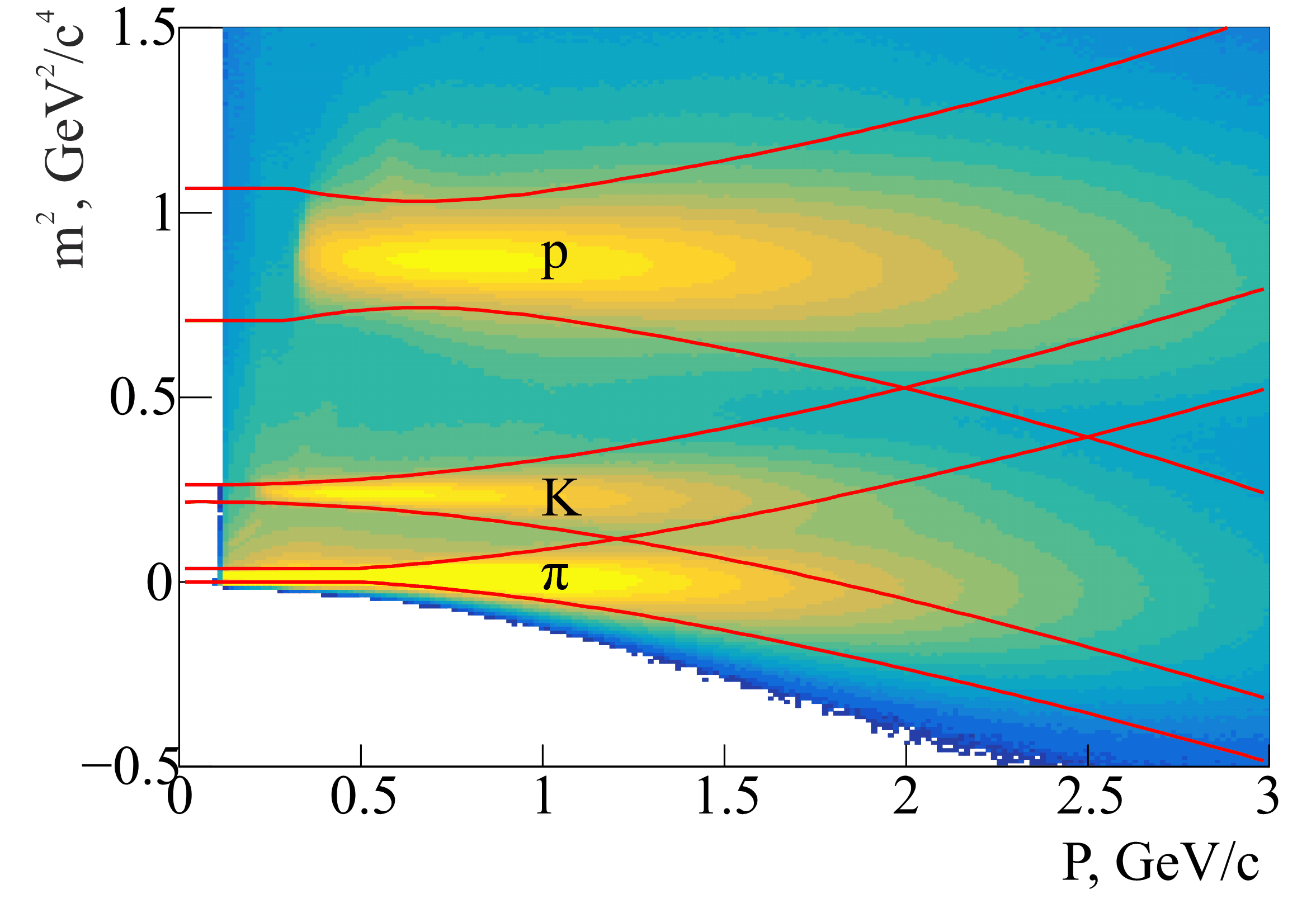}
\includegraphics[width=0.49\linewidth]{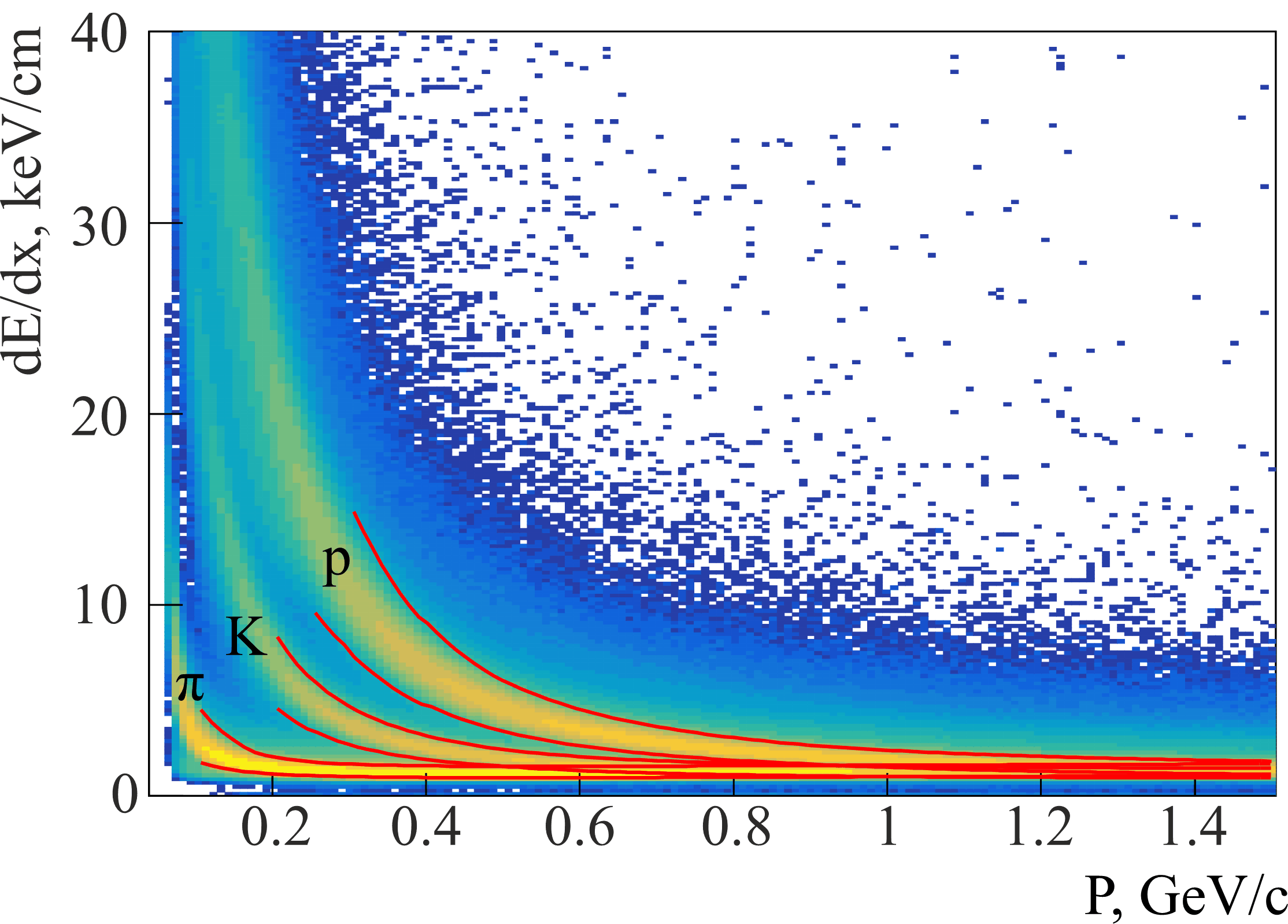}
\vspace{-3mm}
\caption{Momentum distribution of squared mass (left) and energy loss (right) of $\pi, K, p$ from the MPD TOF and TPC systems. Red lines --- 3$\sigma$ borders.}
\end{center}
\labelf{fig01}
\vspace{-5mm}
\end{figure}

In the process of designing the MPD experimental setup, it became necessary to evaluate the properties of the proposed detector subsystems. 
One of the important properties of such setup is particles identification quality. The identification of the particle type in MPD can be carried out using 
several subsystems. TPC and TOF are usually used for this. Using the time-of-flight measurements we can plot the dependence of the squared mass
of particles on its momentum (Fig.~\ref{fig01} on the left). 
On the other hand, we can identify particles by their energy losses $dE/dx$ in the TPC volume using the dependence of energy losses on momentum 
(Fig.~\ref{fig01} on the right). The typical way to separate particles using such a dependencies is a "$n$-sigma" method. The essence of the "$n$-sigma" 
method is that the value by which particles are identified is randomly distributed and has a certain dispersion $\sigma$. The distribution for each type of 
identified particle is approximated over the entire momentum range. As a result, "$n$-sigma" boundaries are determined. Typically, boundaries of 
3$\sigma$ are chosen. Particles are identified by falling within these boundaries.

Identification using the TOF system alone is ineffective for low-momentum particles because they do not reach the TOF detectors due to the magnetic field. 
Therefore, for particles with small momentum it is necessary to use information about the energy loss in the TPC. But, as can be seen from Figure 2, the 
distribution of energy losses for particles with momentum greater than 1 GeV/c strongly overlap. Thus, it is most effective to combine information from both 
systems. When the combination of time-of-flight and energy loss both used to identification, the particles are well separated in a wide range of momenta (Fig.~\ref{fig02}).

\begin{figure}
\begin{center}
\vspace{-3mm}
\includegraphics[width=0.325\linewidth]{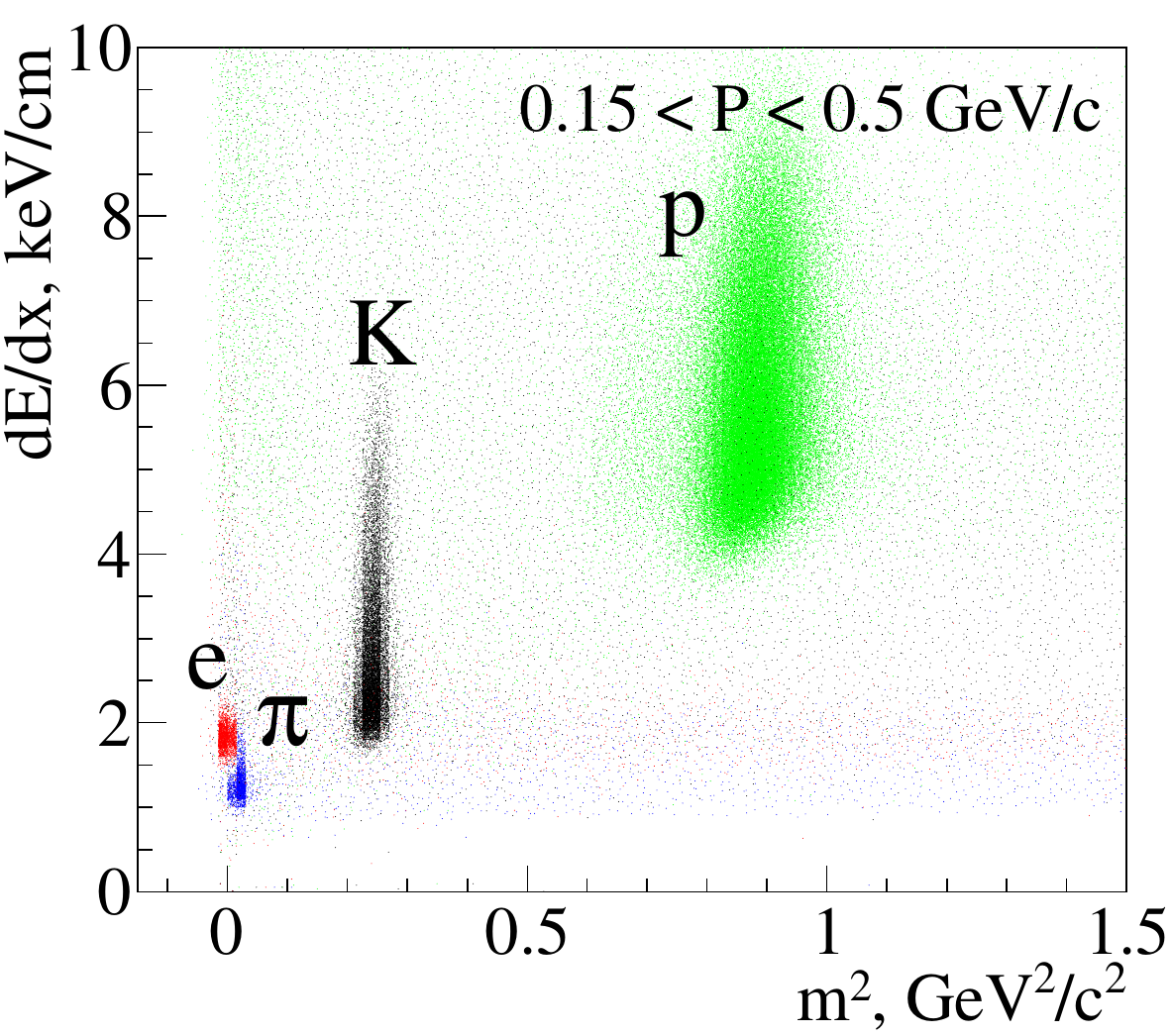}
\includegraphics[width=0.325\linewidth]{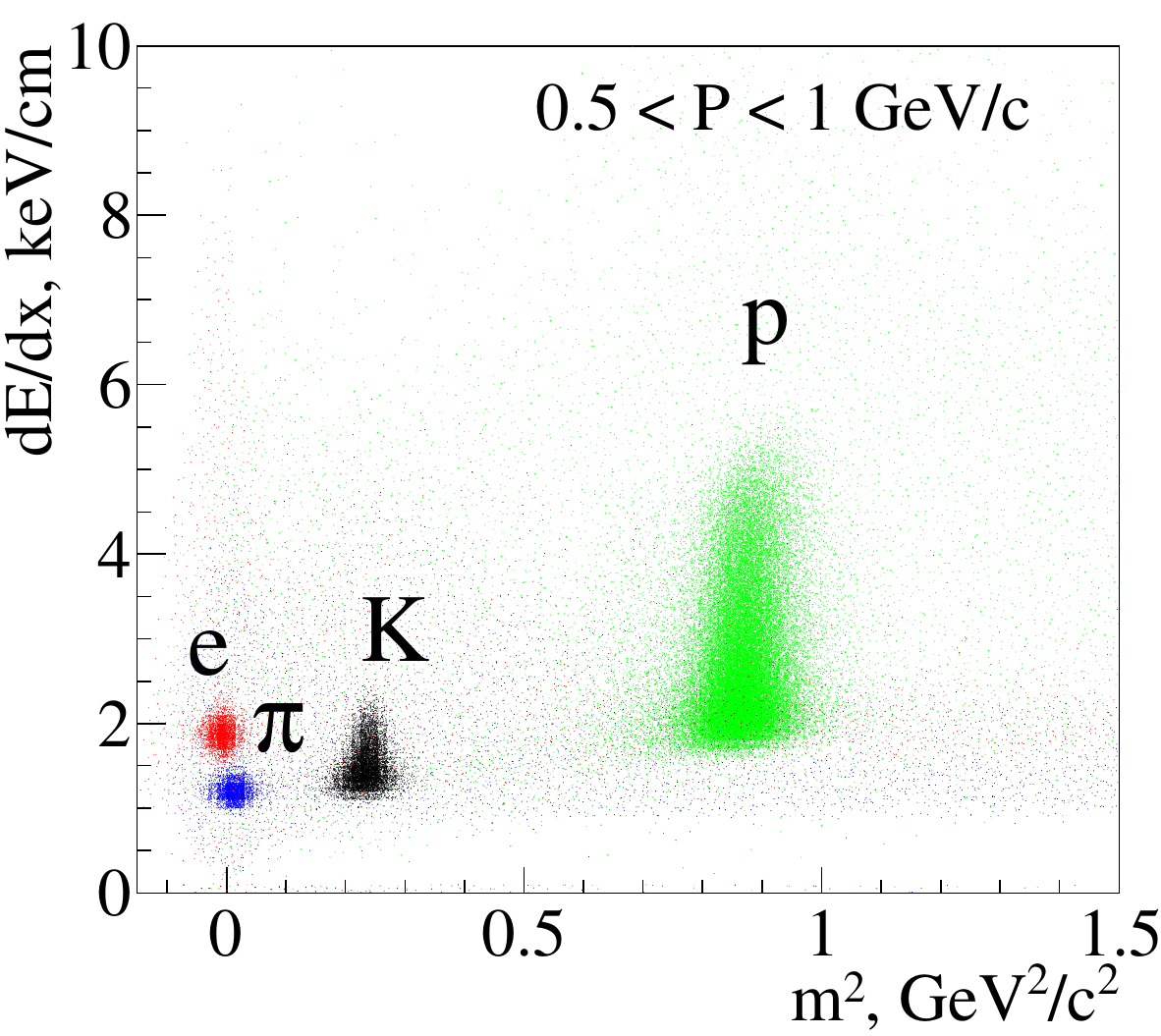}
\includegraphics[width=0.325\linewidth]{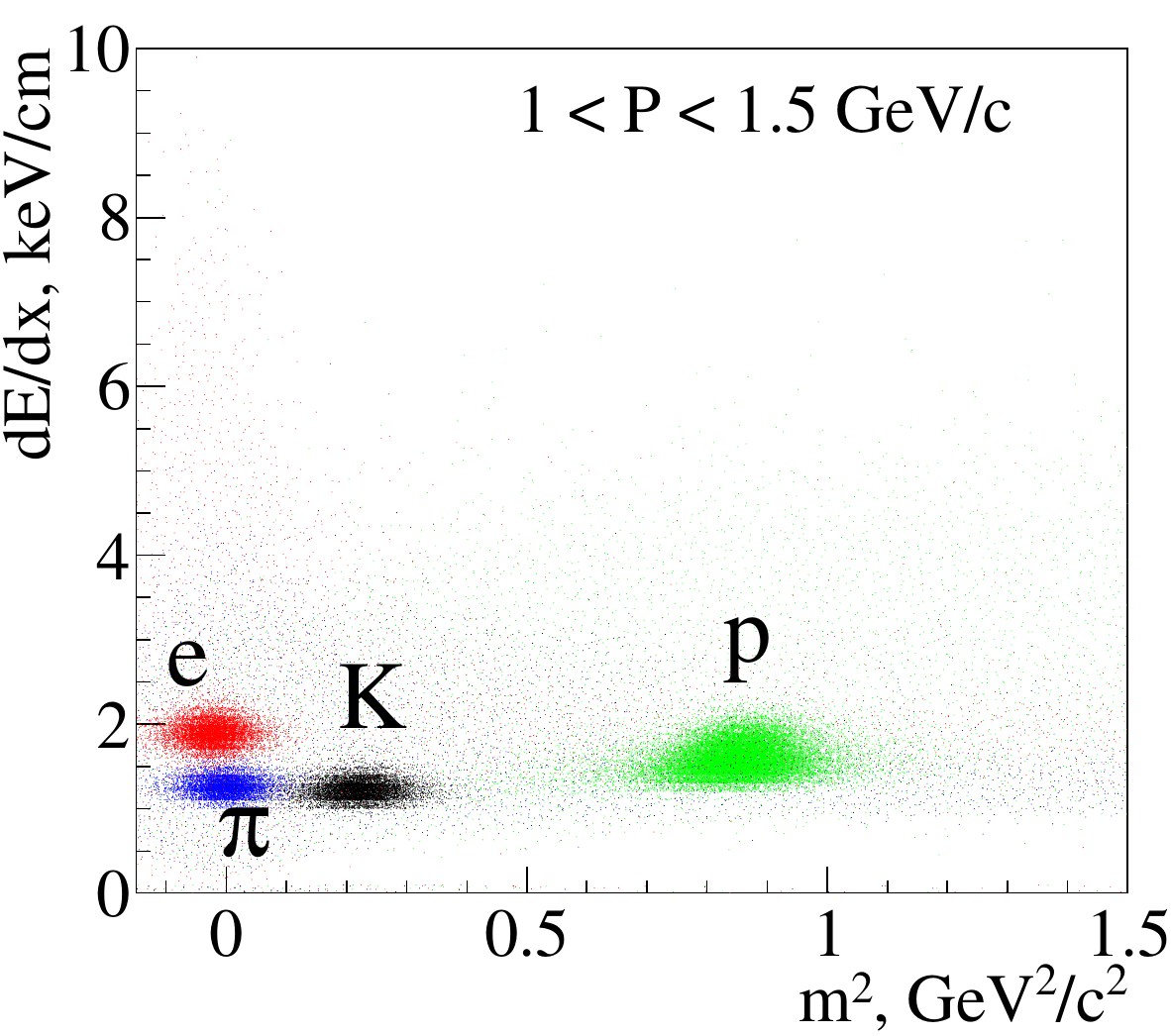}
\vspace{-3mm}
\caption{Combined ToF and $dE/dx$ identification in different momentum range.}
\end{center}
\labelf{fig02}
\vspace{-5mm}
\end{figure}

The combined identification by the TOF and TPC using "$n$-sigma" method provides particles separation with high efficiency and relatively low 
contamination for pions and protons (Fig.~\ref{fig03}). The high level of contamination of $K$-mesons with momenta above 1 GeV/c can be explained 
by the strong overlap of both the squared mass and the energy losses of other particles in this region.

\begin{figure}[b]
\begin{center}
\vspace{-3mm}
\includegraphics[width=1\linewidth]{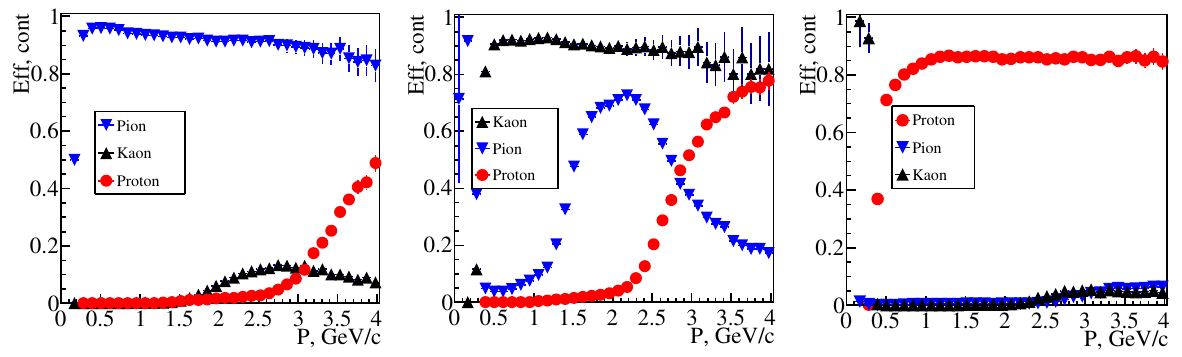}
\vspace{-3mm}
\caption{Efficiency and contamination of $\pi, K, p$ identification using $3\sigma$ method.}
\end{center}
\labelf{fig03}
\vspace{-5mm}
\end{figure}

Such a rather simple and effective approach was used to preliminary evaluation of the possibilities of identification by the MPD 
facility in computer simulation of the physical processes that occur during the collision of ions at the NICA collider. However, when it is necessary to 
separate a certain type of particles with high purity (for example, to study a specific particle decay channel), this method is not effective enough due 
to the high level of contamination. The statistical method described in the next chapter makes it possible to combine data from different detectors 
more effectively to improve the purity of identification.

\label{sec:PID_Bayes}
\section{Principle of applying the Bayesian approach to particle identification}

A more advanced approach to particle identification is to use the statistical Bayesian method \cite{4:ALICE-Bayes}.

As already shown above, the simplest way to identify particles is to select them by TOF or TPC separately or in combination according to a certain 
value $V$ obtained in each subsystem used for identification. For example, $V_{TOF}=m^2$ is the square of mass provided by the time-of-flight 
for a particular particle or $V_{TPC}=dE/dx$ is ionization losses in the TPC gas.

To move to the Bayesian identification approach, it is necessary to introduce the particle detection probability threshold into the data analysis. 
The threshold should functionally depends both on the measured value $V$ and on some other value $V_{exp}$ which can be called the detector 
response. Unlike the measured value $V$, the value $V_{exp}$ is the expected value, which functionally depends both on the properties of detectors 
(time resolution, momentum resolution, etc.) and on the properties of the identified particle (mass, momentum, charge, track length). 
The expected value $V_{exp}$ function or distribution can be obtained by the Monte Carlo method if we know all these properties. If several detectors 
or several different independent values from one detector subsystem are used for identification, then there will be as many such expected value 
distributions as independent experimental values.

In the simplest case, for a detector with a Gaussian response, the function $V_{exp}$ is the mean value of $\hat{V}(H_i)$ with standard deviation 
$\sigma_{i}$, where $H_i$ is the hypothesis of detecting a particle of sort $i$. Thus, for particle $i$, the probability threshold can be represented 
as the number of standard deviations of the value $V$ from the expected mean value $\hat{V}(H_i)$:
\begin{equation}
\label{eq1}
n_{\sigma_{ij}}=\frac{V_j-\hat{V}(H_i)_j}{\sigma_{ij}},
\end{equation}
where $j$ is the detector (TOF, TPC, ...) from which the value $V_j$ is obtained, and $\sigma_{ji}$ is the standard deviation or resolution 
of the expected mean value $\hat{V}(H_i)_j$, which is individual for both detector $j$ and particle $i$. By setting a probability threshold, 
an appropriate decision is made to accept the hypothesis that a given track corresponds to a certain type of particle. For detectors with a 
non-Gaussian response form, one has to find more complex parameterizations that correspond to a specific value received from the detector.

For a normally distributed variable $V$, we can introduce conditional probability that when a particle of sort $i$ passes through detector $j$, 
the value $V$ will be registered in this detector:
\begin{equation}
\label{eq2}
P(V|H_i)=\frac{1}{\sqrt{2\pi}\sigma}e^{-\frac12 n_\sigma^2}=\frac{1}{\sqrt{2\pi}\sigma}e^{-\frac{(V-\hat V(H_{i}))^2}{2\sigma_i^2}}.
\end{equation}

Since we work in terms of probabilities (and the responses of different detectors do not depend on each other), the combined probability for combined 
identification can be obtained by multiplying the probabilities from each detector:
\begin{equation}
\label{eq3}
P(\overrightarrow{V}|H_i)=\prod_{j=TOF, TPC, ...}P_{j}(V_{j}|H_{i}),
\end{equation}
where $\overrightarrow{V}=(V_{TOF},V_{TPC}, ...)$

From the Bayes' theorem, it is possible to express the desired probability that the measured value $V$ in the detector was caused by the particle $H_i$, 
using the conditional probability $P(V|H_i)$ from the Eq.~(\ref{eq2}) and from the $a~priori$ probability for this type of particle:

\begin{equation}
\label{eq4}
P(H_i|V)=\frac{P(V|H_i)C(H_i)}{\sum_{k=\pi,K,p,e...}P(V|H_k)C(H_k)}.
\end{equation}

The $P(H_i|V)$ is $a~posteriori$ probability of the hypothesis that the appearance of the value $V$ in the detector was caused by a particle of type $i$. 
$P(V|H_i)$ is the Monte Carlo probability of the appearance of the value $V$ for a particle of type $i$, and $C(H_i)$ is $a~priori$ probability of passing  
a particle of type $i$ through the detector, which is called the "prior". Priors can be called "the most precise assumption" about the true yield of particles 
per event. In the denominator of the Eq.~(\ref{eq4}) is the normalization coefficient, which is the sum of the numerators for all hypotheses under 
consideration (types of particles). The main property of Eq.~(\ref{eq4}) is that can set a criterion for the purity of particle identification by changing the 
level of the probability threshold $P(H_i|V)$. The purity of identification can be defined as the ratio of the number of correctly identified tracks to the total 
number of selected ones. 

The quality of identification using the Bayesian approach depends on the choice of initial $a~priori$ probabilities for each particle. This article describes an 
iterative method for finding priors. For the first iteration, the values of all priors can be set equal to one in the entire momentum region. 
Further, according to the probabilities $P(H_i|V)$ calculated from Eq.~(\ref{eq4}), it is possible to obtain the output spectra of type $i$ particles 
depending on the momentum $p$ which after normalization can be taken as priors $C_{n+1}(H_i)$ for the next iteration:

\begin{equation}
\label{eq5}
Y(H_i|p)=\sum_{\rm{All~detectors}}P_n(H_i|V)=C_{n+1}(H_i).
\end{equation}

By repeating this procedure iteratively, at a certain iteration, it is possible to achieve convergence of priors to a certain limit value, at which the priors 
practically cease to change with an increase in the iteration number. As a result, the best fit of priors to the available set of experimental or reconstructed 
data is achieved and the results of data processing are independent from the initially selected priors. It should be noted that the convergence of priors 
depends on the resolution of the detectors. The better the quality of measurements from the detector, the faster (in fewer iterations) the priors converge.

\label{sec:PID_eff_cont}
\section{Calculation of identification efficiency and contamination}

The quality of any method of identification can be evaluated by calculating the efficiency of identifying a certain type of particles and their contamination 
with other particles. Identification efficiency is the ratio of the number of correctly identified particles of a certain type to their actual yield under given 
conditions:

\begin{equation}
\label{eq6}
\varepsilon_i=\frac{N_{i~identified~as~i}}{N_{i~true}}.
\end{equation}

Contamination of particles of type $i$ by other particles is the ratio of the sum of all particles $j$ that are identified as $i$, but are not $i$, to all particles 
that were identified as $i$:

\begin{equation}
\label{eq7}
c_{ij}=\frac{N_{j~identified~as~i}}{N_{i~measured}}, i\ne j.
\end{equation}

At the same time, it is important to distinguish contamination particles from incorrectly identified ones. The probability of misidentified particles is the ratio 
of the number of real (true) particles of sort $i$, which are identified as particles of a different sort, to the total yield of particles of sort $i$.

%
%

\label{sec:PID_results}
\section{Results of the Bayesian approach for identification in MPDRoot}

To integrate the developed identification algorithm, a program code was written that allows, within the MPDRoot \cite{5:MPDRoot} framework, to apply 
the proposed method to both real experimental data and realistically reconstructed collisions of heavy ions in the MPD experiment. 

A combined space of values $V$ for TOF and TPC detectors was considered in this algorithm: $P$~(GeV/c), $m^2$~(GeV$^{2}$/c$^{4}$), $dE/dx$~(keV/cm). 
Four types of particles were selected for identification: $e^{\pm}, \pi^{\pm}, K^{\pm}, p(\bar{p})$. Due to the number of values $V$ equal to three, it became 
possible to implement the container of the probability matrix in the form of a TH3D Root class. The TH3D class optimizes (compresses) highly sparse (with a large 
number of empty "bins") arrays well.

Calculations are performed as follows. First, the macro for events reconstruction is running on sufficient statistic of the corresponding physical Monte Carlo data. 
Probability $P(V|H_{i})$ and priors $C(H_{i}$ matrices  based on the first iteration running are created and stored as TH3D objects in $TRoot$ files. Then the 
same macro is running with loaded from files  $P(V|H_{i})$ matrices and priors $C(H_i)$ from the previous stage. This iteration is repeated until the priors converge.

For the first iteration, input priors are automatically created with uniform ($C(H_i ) = 1$) probability density functions (PDF) for all types of particles. 
The second iteration of calculations can be performed both on Monte Carlo data and on experimental data.

Proposed method was applied to the $3*10^5$ minimum bias LAQGSM events for collisions of Au+Au ions with energy $\sqrt{S_{NN}} =$ 11.5 GeV 
in the MPD setup. The following constraints were used for matching of reconstructed in TPC tracks with hits in TOF. Tracks are selected only for particles 
from primary interaction vertices and all tracks must have at least 20 hits in TPC.

\begin{figure}
\begin{center}
\vspace{-0mm}
\includegraphics[width=1\linewidth]{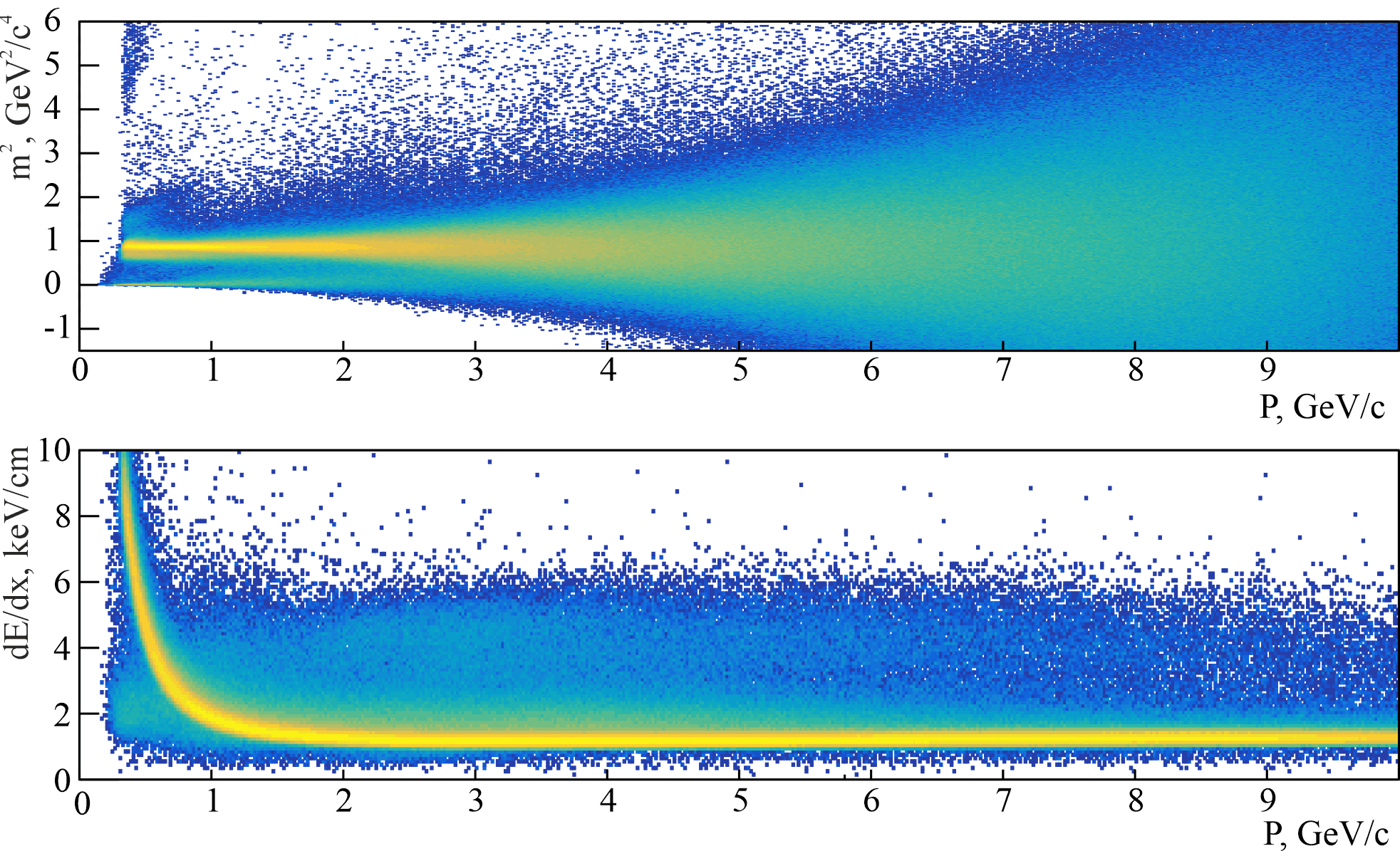}
\vspace{-3mm}
\caption{Monte Carlo PDF for proton on a Box generator. The squared mass $m^2$ (TOF) is at the top, the energy losses $dE/dx$ (TPC) is at the bottom.}
\end{center}
\labelf{fig04}
\vspace{-3mm}
\end{figure}

As it was described above, for the first iteration we need to fill probabilities $P(V|H_{i})$ and priors matrices $C(H_i)$. All values of the priors matrices 
are initially equal to one by default. To obtain expected values  $\hat{V}(H_i)$ and standard deviation $\sigma_{i}$ which are needed for filling probability 
matrices $P(V|H_i)$ a simple Box event generator was used. It provides uniform distribution of spectra of particles over the momentum $p$ and the angles 
of escaping of particles $45^{\circ}<\theta<135^{\circ}$. Thus, a statistically equivalent expected values are provided for any momenta of particles. 
Fig.~\ref{fig04} shows, as an example, the distributions of the squared mass and the ionization losses of the proton in dependence on the momentum 
obtained as a result of the calculations described above. These $P(V|H_{i})$ distributions can be considered as probability density functions (PDF) for each 
type of particle.

\begin{figure}
\begin{center}
\vspace{-3mm}
\includegraphics[width=0.49\linewidth]{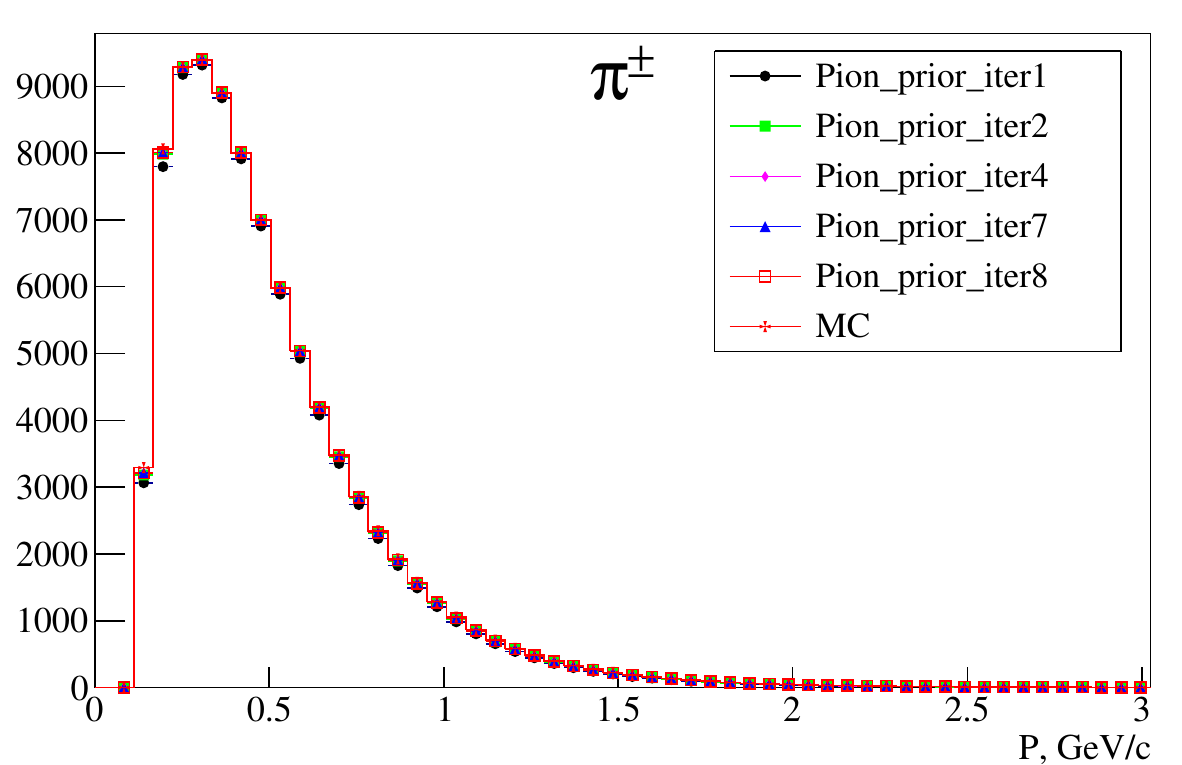}
\includegraphics[width=0.49\linewidth]{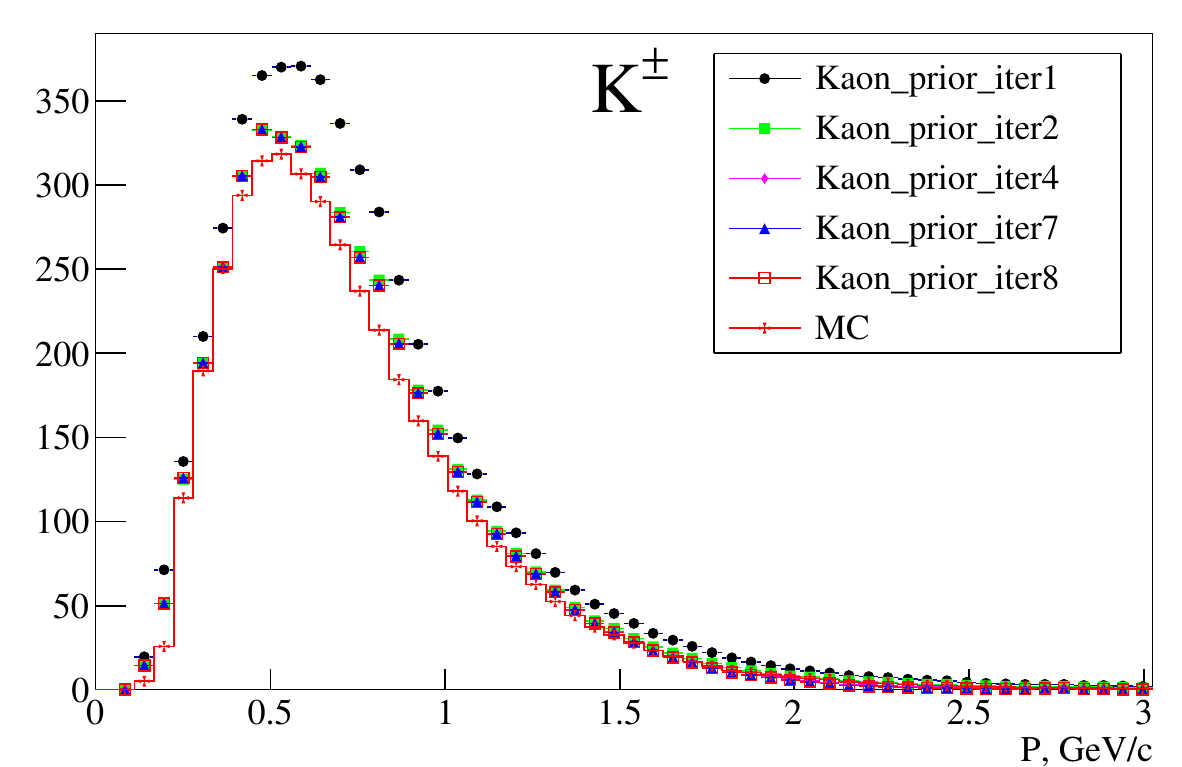}
\includegraphics[width=0.49\linewidth]{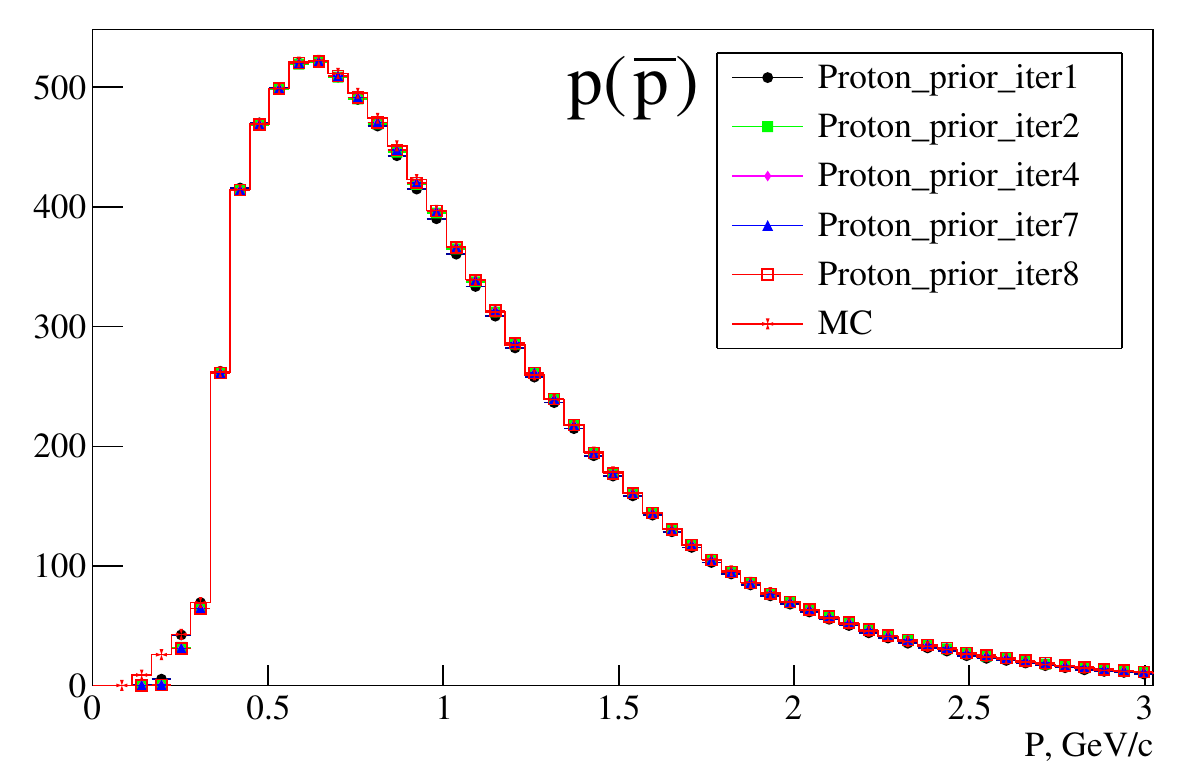}
\includegraphics[width=0.49\linewidth]{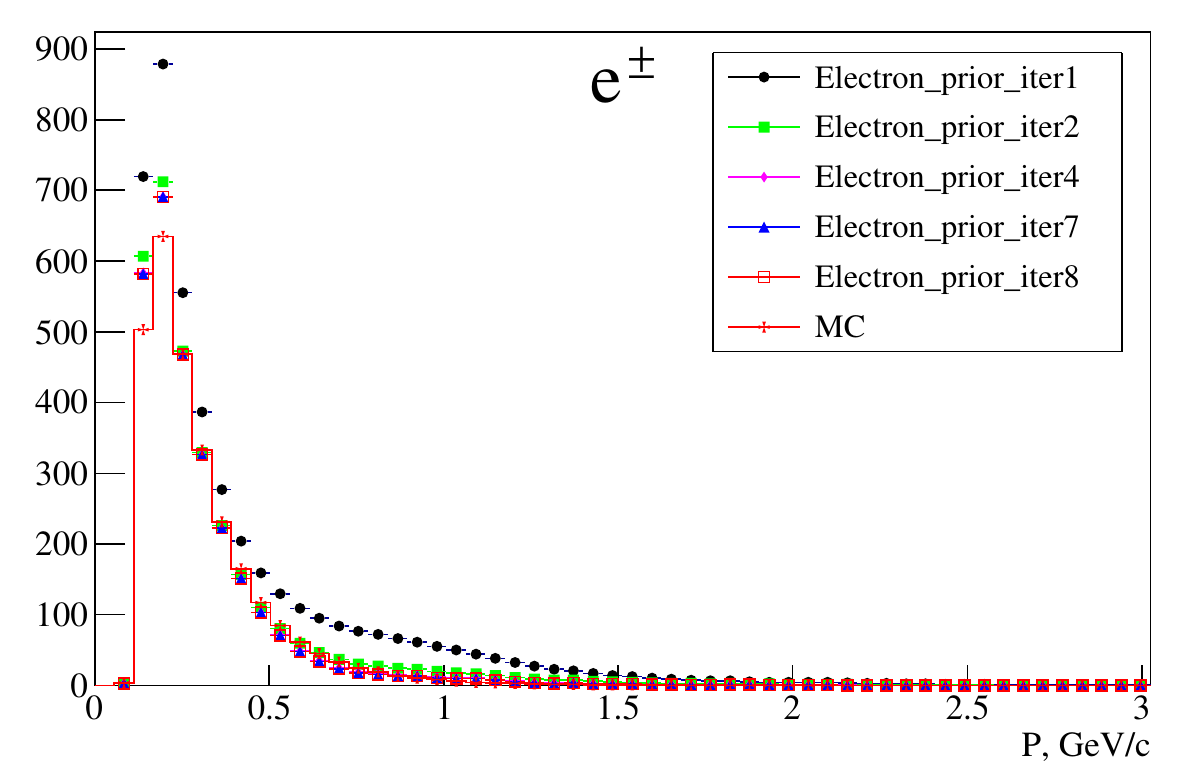}
\vspace{-3mm}
\caption{Evolution of priors for $\pi^{\pm}$-meson, $K^{\pm}$-meson, (anti-) proton, and electron (positron). Red lines corresponds to real (MC) yields of particles.}
\end{center}
\labelf{fig05}
\vspace{-5mm}
\end{figure}

The output spectra of the identified particles that we use as priors are shown in Fig.~\ref{fig05}. It can be seen that for proton and pion, 
due to their large physical statistics, $a~priori$ probabilities converge in 2-3 iterations and further calculations practically do not change the shape of 
the distribution. Priors for kaons and electrons converge slowly. The prior for kaon converges on the third or fourth 
iteration.The prior for electrons converge slower than all identifiable particles due to low statistics and strong overlap with $\pi$ mesons.

\begin{figure}[b]
\begin{center}
\vspace{-3mm}
\includegraphics[width=0.49\linewidth]{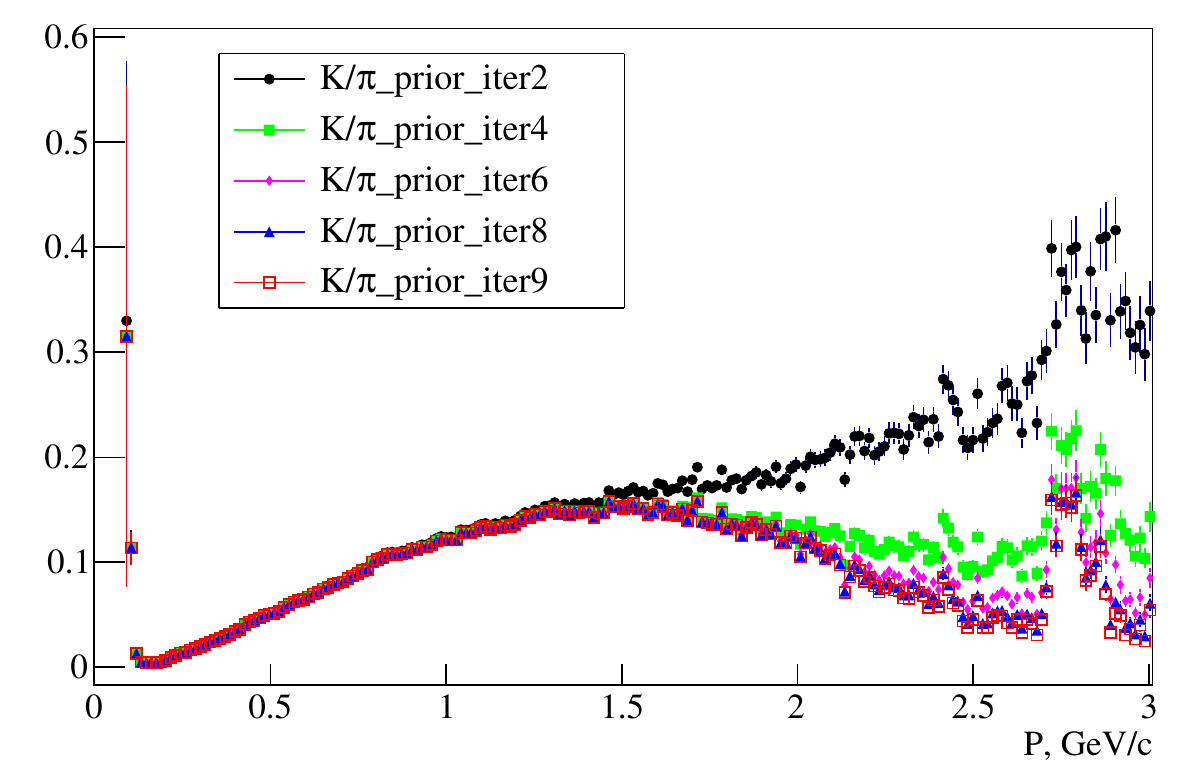}
\includegraphics[width=0.49\linewidth]{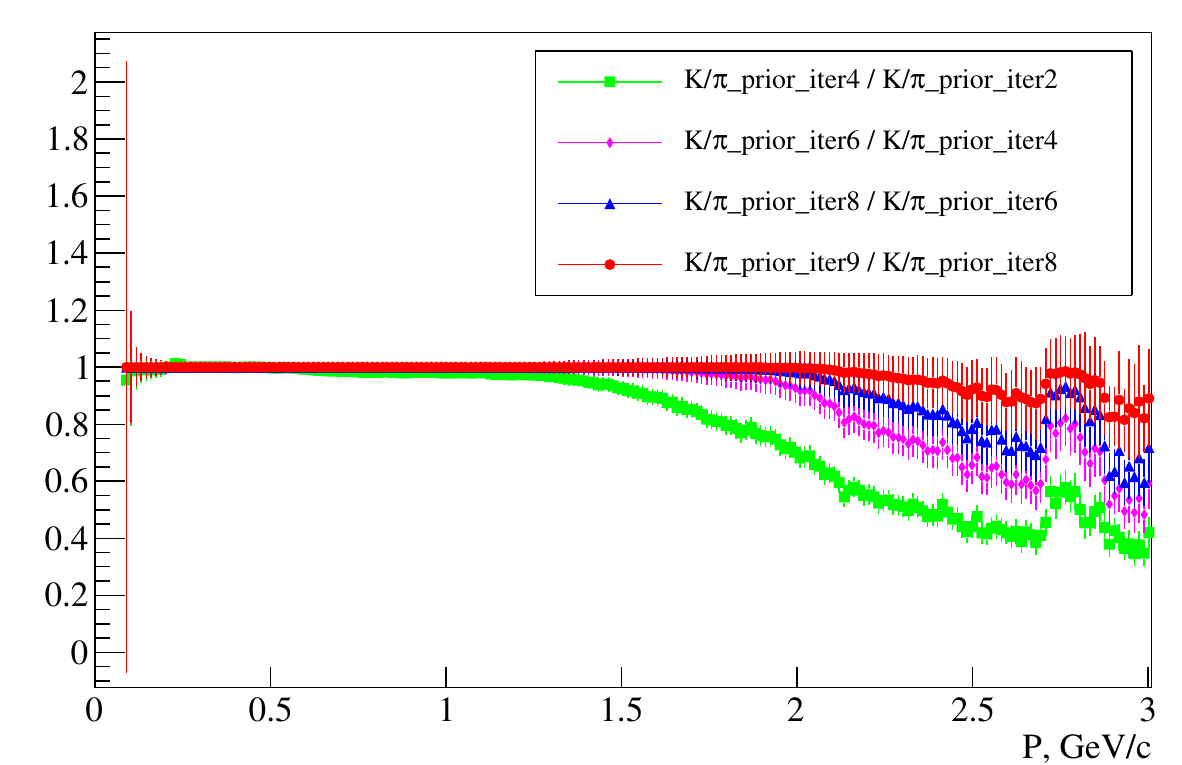}
\vspace{-3mm}
\caption{Evolution of ratios of priors  $K/\pi$ (left) and ratios of priors between different steps $(K/\pi)_{step~n}/(K/\pi)_{step~n-1}$ (right).}
\end{center}
\labelf{fig06}
\vspace{-5mm}
\end{figure}

\begin{figure}
\begin{center}
\vspace{-3mm}
\includegraphics[width=0.49\linewidth]{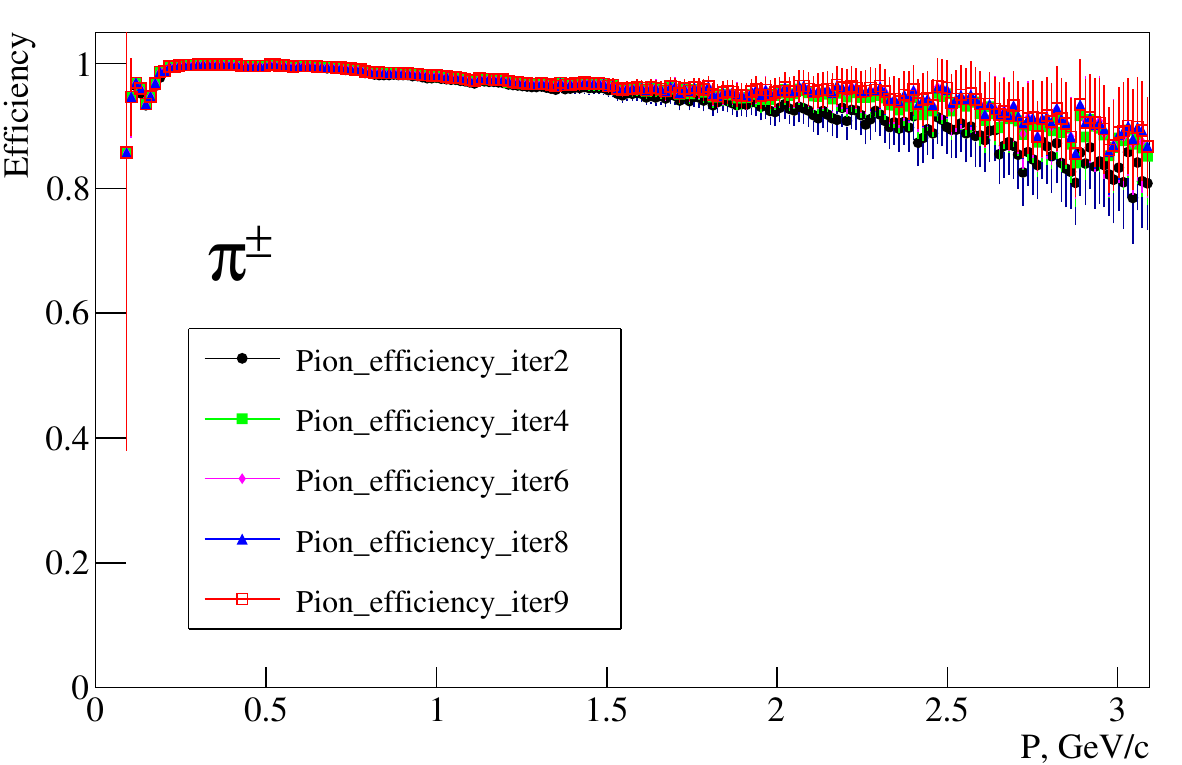}
\includegraphics[width=0.49\linewidth]{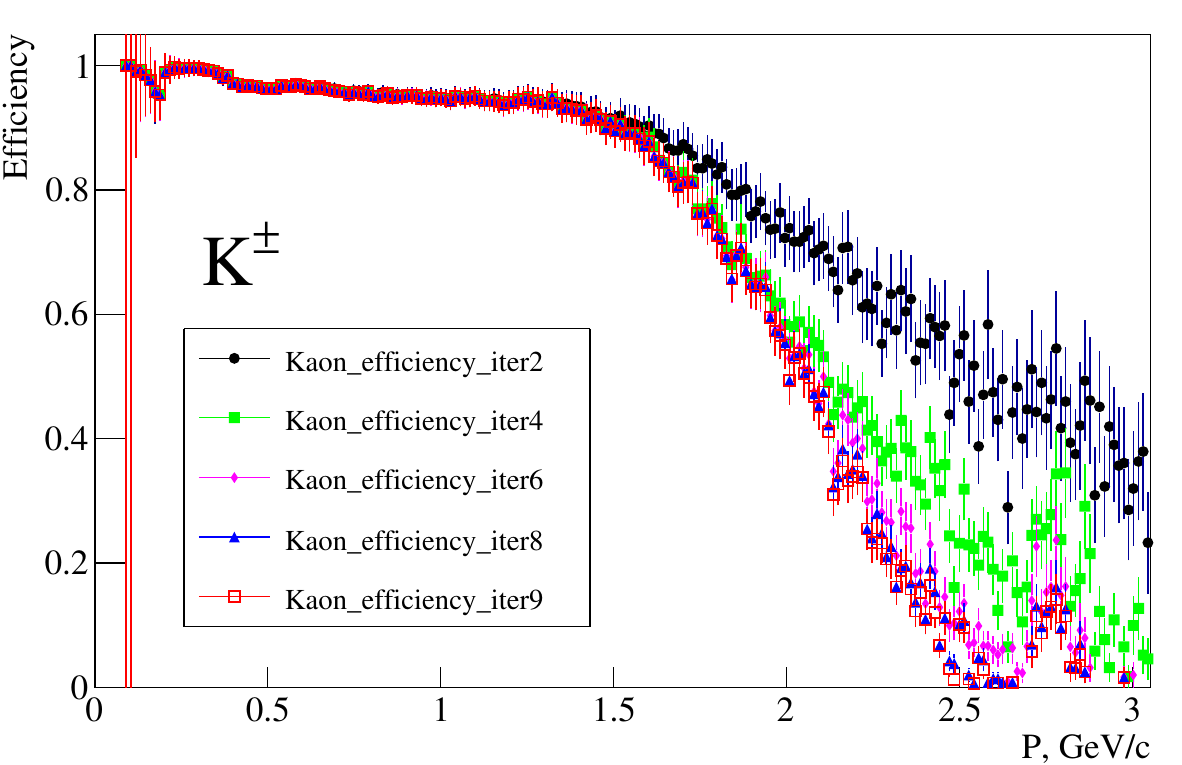}
\vspace{-3mm}
\caption{Evolution of identification efficiency for $\pi^{\pm}$ (left) and $K^{\pm}$ (right).}
\end{center}
\labelf{fig07}
\vspace{-5mm}
\end{figure}

The convergence of priors can be best estimated by studying the ratio of priors $K/\pi$ (Fig.~\ref{fig06} on the left). The ratio of priors converges on 
the second-third iteration in the region of momentum up to 1.5~GeV/c. But for higher momenta even six iterations are not enough. 

This is especially clearly seen in Fig.~\ref{fig06} on the right, which shows the ratios of previous and following iterations of priors $K/\pi$ . 
The criterion of convergence of $K/\pi$ ratios can be considered the best way to evaluate the convergence of priors for individual particles.

As shown in the Fig.~\ref{fig07}, the efficiency of identification of pions improves when using the iterative Bayesian approach with each  
iteration. The efficiency of kaons identification unlike pions decreases for the momentum region > 1.5 GeV/c. This is due to the fact that with 
each subsequent iteration the priors converge  to the real yields of particles and a larger number of kaons do not fall under the selected 
probability threshold. On the other hand, better identification purity is ensured.

Identification of particles by probabilities $P(H_i|V)$ for comparison was performed  in this work according to two following methods:
\begin{enumerate}
\item The maximum probability method. The track belongs to the type of particles for which the probability is maximum.
\item Fixed threshold method. The track is considered to be the type of particle for which the probability is higher than the threshold. Obviously, at a higher 
probability threshold, identification efficiency decreases, but purity improves. This is due to the fact that at a threshold of <0.5 there are cases when one 
track can correspond to two types of particles, and at <0.333 -- to three.
\end{enumerate}

There is also a weighted method, which didn't used in this work. It is used to select only specific particles, such as those involved in the decay of the primary 
particle. The weight in this case is defined as the product of Bayesian probabilities for the allocated particles. This method most clearly identifies the required 
particle decay channel. 

\begin{figure}
\begin{center}
\vspace{-3mm}
\includegraphics[width=1\linewidth]{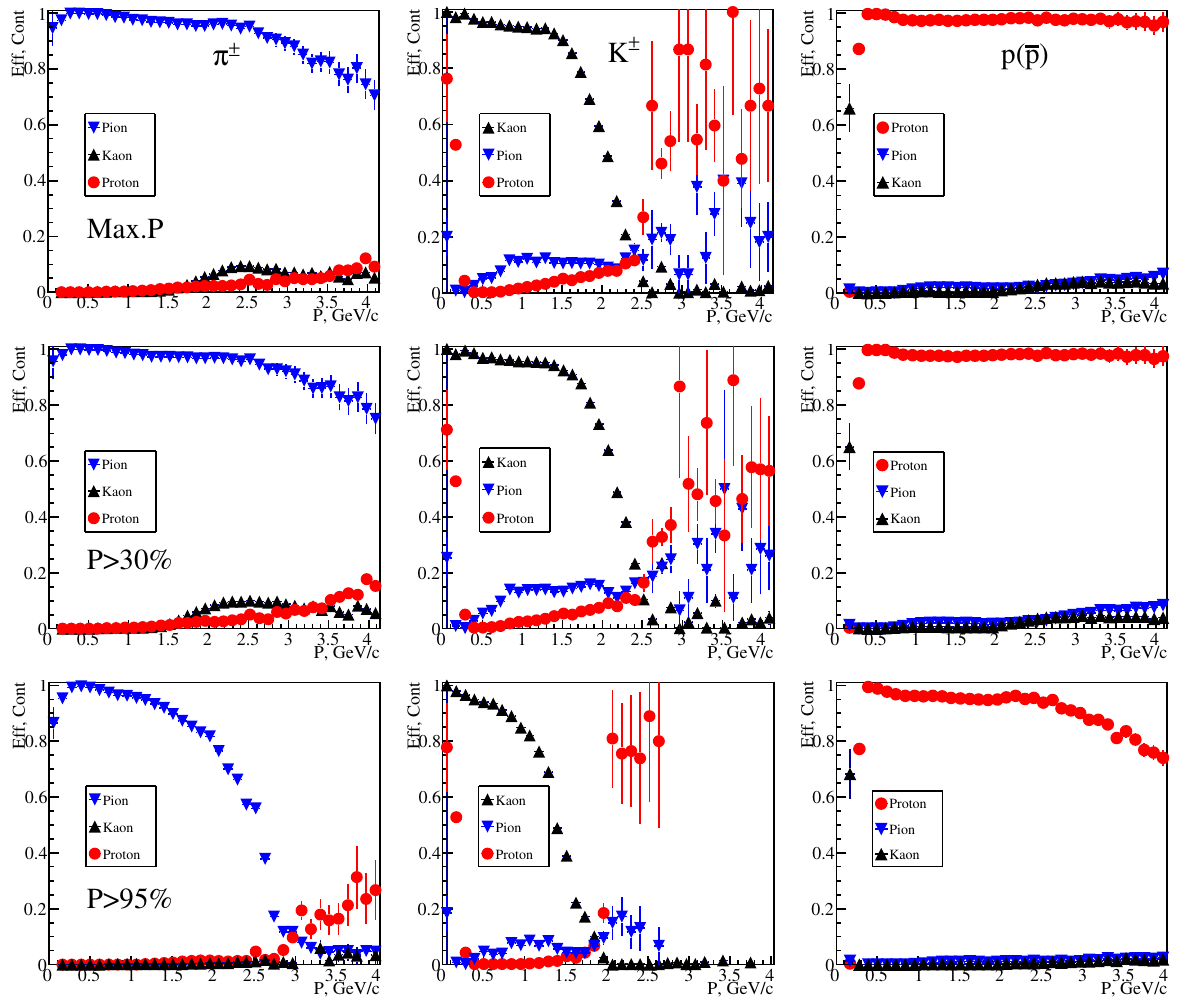}
\vspace{-3mm}
\caption{Identification efficiency and contamination for pion (left column), kaon (middle column) and proton (right column) by various methods: maximum 
probability (top raw), fixed threshold of 30\% (middle raw) and 95\% (bottom raw).}
\end{center}
\labelf{fig08}
\vspace{-5mm}
\end{figure}

Fig.~\ref{fig08} show the efficiency and contamination of proton, pion and kaon identificationafter applying 10 iterations of the Bayesian approach. 
When comparing particle selection methods, we can see that the maximum probability method is almost equivalent to the method with a fixed 
threshold of 30\%. Increasing the threshold to 95\% worsens the identification efficiency for particles with large momentum, but on the other hand, 
noticeably reduces contaminations. This property of the proposed method is very important in the study of processes with a large background, when 
statistics are sacrificed to improve the signal-to-noise ratio. Low identification efficiency at high probability thresholds does not affect the determination 
of total particle yields. To determine the yields, it is necessary to normalize the experimental particle yields to efficiency distributions obtained from 
Monte Carlo simulations.

\label{sec:Concl}
\section{Conclusions}

A preliminary evaluation of the identification method using the Bayesian approach showed its greater flexibility compared to other methods used for 
identification. The use of this approach does not always increase the efficiency of identification, but at the same time allows it to be carried out clearly. 
At the same time, it should be noted that the quality of identification strongly depends on the method of selecting the initial priors. In this work, the 
priors are defined most closely to a realistic data analysis. That is, we believe that we don't know anything about the actual particle yields. Additionally, 
the real characteristics of the detectors and the presence of a background of secondary particles are taken into account. Because of this, the results 
may look worse than in previous publications \cite{3:TOF-MPD}. But at the same time they are more realistic.

In any case, the real quality of the various identification approaches will be revealed only when experimental physical data are obtained in the MPD 
experiment with collisions of ions at the NICA collider.

\label{sec:ackn}
\section*{Acknowledgements}

The work was financially supported by a Program of the Ministry of Education and Science of the Russian Federation for higher education establishments, 
project No.FZWG-2020-0032 (2019-1569). This research was supported by the Russian Foundation for Basic Research: grant 18-02-40051.

\end{document}